\documentclass[twocolumn,superscriptaddress,PRL]{revtex4-1}
\usepackage{graphicx}
\usepackage{dcolumn}
\usepackage{bm}

\begin{document}


\title{
Current-driven instability of quantum anomalous Hall effect \\
in ferromagnetic topological insulators
}

\author{Minoru Kawamura}
\email{minoru@riken.jp}
\affiliation{ 
RIKEN Center for Emergent Matter Science (CEMS), Wako 351-0198, Japan}

\author{Ryutaro Yoshimi}
\affiliation{ 
RIKEN Center for Emergent Matter Science (CEMS), Wako 351-0198, Japan}

\author{Atsushi Tsukazaki}
\affiliation{
Institute for Materials Research, Tohoku University, Sendai 980-8577, Japan}

\author{Kei~S.~Takahashi}
\affiliation{ 
RIKEN Center for Emergent Matter Science (CEMS), Wako 351-0198, Japan}
\affiliation{
PRESTO, Japan Science and Technology Agency (JST), Chiyoda-ku, Tokyo 102-0075, Japan}

\author{Masashi Kawasaki}
\affiliation{ 
RIKEN Center for Emergent Matter Science (CEMS), Wako 351-0198, Japan}
\affiliation{ 
Department of Applied Physics and Quantum-phase Electronics Center (QPEC),
University of Tokyo, Tokyo 113-8656, Japan}

\author{Yoshinori Tokura}
\affiliation{ 
RIKEN Center for Emergent Matter Science (CEMS), Wako 351-0198, Japan}
\affiliation{ 
Department of Applied Physics and Quantum-phase Electronics Center (QPEC),
University of Tokyo, Tokyo 113-8656, Japan}

\date{\today}

\begin{abstract}
Instability of quantum anomalous Hall (QAH) effect has been studied as functions
of electric current and temperature in ferromagnetic topological insulator thin films.
We find that a characteristic current for the breakdown of the QAH effect
is roughly proportional to the Hall-bar width,
indicating that  Hall electric field is relevant to the breakdown.
We also find 
that electron transport is dominated 
by variable range hopping (VRH) at low temperatures.
Combining the current and temperature dependences of the conductivity
in the VRH regime,
the localization length of the QAH state is evaluated to be about  5~$\mu$m.
The long localization length suggests  a marginally insulating nature of
the QAH state due to a large number of in-gap states.
\end{abstract}

\maketitle

Electronic phenomena arising from 
topological nature of  band structures in solids
has been gaining great interests in condensed-matter physics,
triggered by the recent discoveries of two-dimensional\cite{Kane, Bernevig, Konig}
and three-dimensional (3D) topological insulators (TIs)\cite{Fu, Hsieh, Brune, Hasan, Qi}.
A 3D TI is an insulator with an energy gap in the band structure of its interior,
while the gap closes at the surface due to the exotic topology 
of the band structure\cite{Fu, Hsieh, Brune, Hasan, Qi}.
One of the most decisive examples of the topology-related phenomena in a 3D TI
is the quantum anomalous Hall effect (QAHE)
found in ferromagnetic thin films\cite{Yu, Nomura, Zhang, Chang1}.
In the quantum anomalous Hall (QAH) state,
the Hall resistance is quantized to $h/e^2$
with vanishing longitudinal resistance, 
similarly to the quantum Hall effect (QHE)\cite{Klitzing}
but without external magnetic field.
The  QAHE has been observed experimentally 
in thin films of  Cr-doped\cite{Chang1, Checkelsky,
Kou, Bestwick, Mogi} 
or V-doped \cite{Chang2, Chang3, Grauer} (Bi, Sb)$_2$Te$_3$
grown by molecular beam epitaxy.
The films show robust ferromagnetism with an easy-axis anisotropy
perpendicular to the films with typical ferromagnetic transition temperatures at around 10 K.
The Hall resistance quantization and the vanishing longitudinal resistance
have been confirmed by transport measurements at low temperatures.

The QAHE can be understood 
as a consequence of an energy gap opening on the topological surface states
by the exchange interaction between the surface electrons
and the magnetic moments\cite{Yu, Nomura, Zhang, Chang1}.
In a thin film form of a ferromagnetic 3D TI with perpendicular magnetic anisotropy, 
the top and bottom surfaces are gapped
while the side surfaces parallel to the magnetization remain gap-less.
On these gap-less side surfaces, one-dimensional chiral edge channels are formed
with a help of quantum confinement effect in a thin-film form\cite{Pertsova}.
When the Fermi energy $E_{\rm F}$ is tuned in the magnetization-induced gap,
the electric current is carried by the chiral edge channels,
resulting in the Hall resistance quantization.
Because the relevant energy gap is induced by the magnetization,
the QAHE can survive even in the absence of an external magnetic field.
The QAHE is believed to be robust against disorder or any other perturbations 
unless the magnetization-induced gap of the surface states is closed.

Despite of the different origins for the energy gap formation;
Landau level splitting in the QHE and exchange gap of surface states in the QAHE,
the QAHE resembles the QHE in many aspects\cite{Checkelsky, Kou, Chang2},
including the preciseness of the  quantized Hall resistance.
The chiral nature of the edge channels, 
prohibiting  back scattering of electrons, 
ensures the precise quantization of the  Hall resistance to $h/e^2$.
Therefore the use of the QAHE for a resistance standard
 is one of the promising applications.
Because the Hall resistance quantization can be achieved
in the absence of an external magnetic field,
realization of the QAHE-based resistance standard
is appealing from a practical viewpoint.

Recent experimental studies concerning the precision
of the quantized resistance
report an error of about $10^{-9}$
in the case of QHE in graphene and GaAs-based systems
\cite{Jeckelmann, Tzalenchuk, Lafont},
but only $10^{-4}$ in the case of QAHE\cite{Chang3, Bestwick}.
The precise quantization of the QAHE 
was achieved only at low temperatures below 100~mK.
The low observable temperature of the QAHE has been
an issue of discussions\cite{Mogi, Li}.
In addition, stability of the QAH state against  electric current matters
for the resistance standard application because a large current
in the order of  ten~$\mu$A  is usually employed
for the high-precision measurements\cite{Jeckelmann, Tzalenchuk, Lafont}.
Although poor stability against electric current
is intuitively expected from the low observable temperature,
direct measurement of the current stability
and comparison with QHE
would provide quantitative understandings of QAHE
toward realization of the zero-field resistance standard.

In this Letter, 
we report instability of the QAHE as functions of electric current and temperature
 in ferromagnetic 3D TI thin films.
The QAHE is broken
when the current applied to the QAH conductor exceeds a characteristic value.
The characteristic current
is roughly proportional to the Hall-bar width,
showing relevance of Hall electric field to the breakdown.
We also find that variable range hopping (VRH)
among localized states contributes to 
the electron transport at low temperatures.
Combining the current and temperature dependences 
of the longitudinal conductivity in the VRH regime,
the localization length of the QAH state is evaluated to be about 5 $\mu$m.
The long localization length as well as  the small characteristic current
for the QAHE breakdown indicates  presence of a large number of in-gap states.

\begin{figure}[tbp]
\centering
\includegraphics[width=8.4cm]{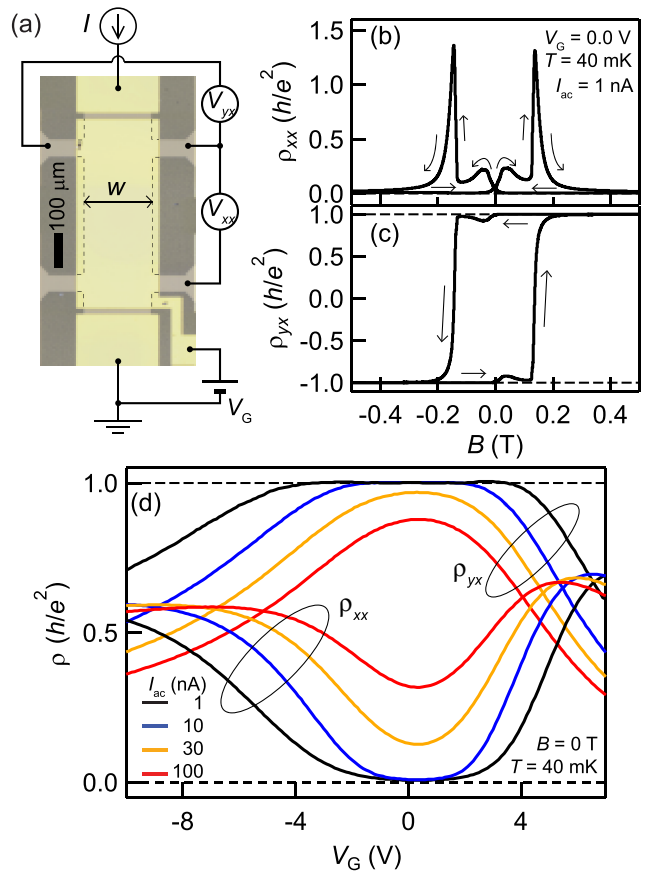}
\caption{\label{fig1} (color online) 
	(a) Photograph of a 200-$\mu$m-wide  Hall bar
	made of ferromagnetic topological insulator 
	${\rm Cr}_{0.1}({\rm Bi}_{0.14}{\rm Sb}_{0.78})_{1.9}{\rm Te}_3$
	covered with Ti/Au top-gate electrode including
	schematic of measurement circuit.
	(b) (c) Magnetic field dependence of the longitudinal resistivity $\rho_{xx}$ (b)
	and the Hall resistivity $\rho_{yx}$ (c) 
	for the 200-$\mu$m-wide Hall bar at $T$ = 40~mK
	with an excitation current $I_{\rm ac}$ =  1~nA at a frequency of 3~Hz.
	Magnetic field was scanned in the positive and negative directions
	at a rate of 0.035~T/min. The arrows indicate the field-scan directions.
	Increase in $\rho_{xx}$ and deviation of $\rho_{yx}$
	from $h/e^2$ around $\pm$0.1~T
	are associated with the temperature increase upon the magnetic-field polarity reversal 
	during the field scan.
	The gate voltage was tuned to the charge neutrality point (CNP) at $V_{\rm G}$ = 0.0~V.
	(d) Gate voltage dependence of $\rho_{xx}$ and $\rho_{yx}$
	for the 200-$\mu$m-wide Hall bar
	at $T$ = 40~mK and $B$ = 0~T after a magnetic training. 
	Traces measured at $I_{\rm ac}$  =  1, 10, 30 and 100~nA are shown.
 }
\end{figure}

Experiments were conducted using
thin  films  of  ferromagnetic TI Cr$_{x}$(Bi$_{1-y}$Sb$_{y}$)$_{2-x}$Te$_3$ grown
on InP (111)A surface  by  molecular  beam  epitaxy.
The  nominal  compositions and the film thickness were $x = 0.1$, $y = 0.78$,
and  9 nm, respectively.
The magnetic ions Cr were doped uniformly throughout the film.
Details of the growth are  described in Ref. \cite{Checkelsky}.
The films were patterned into a Hall-bar shape
using photo lithography and chemical wet etching
followed by formation of Ti/Au contact electrodes by electron-beam evaporation.
Three Hall bars were prepared with different widths 
$W$ = 50, 100, and 200 $\mu$m [Fig. 1(a)].
To tune the Fermi energy,
top gate electrodes (Ti/Au) were formed on the top of the Hall bars
after growing aluminum oxide by atomic layer deposition as a gate dielectric.
Transport measurements were conducted
using standard dc or low-frequency ac lock-in methods
 in a dilution refrigerator with a base temperature $T$ = 40~mK.
The longitudinal $\rho_{xx}$ and the Hall resistivity $\rho_{yx}$
were calculated 
by dividing the measured resistances
by the aspect ratio of the Hall bars.

The transport properties of the Hall bars clearly exhibit the QAHE 
as shown in Fig. 1 (b)-(d).
When external magnetic field $B$ normal to the film
was scanned in the positive and negative directions at $T$ = 40~mK,
$\rho_{yx}$-$B$ curves make a sharp square hysteresis loop,
reflecting changes in the magnetization direction of the film[Fig. 1(c)].
The saturated values of $\rho_{yx}$
is close to the quantized resistance $h/e^2$.
The magnetization reversals are accompanied by resistivity peaks
in the $\rho_{xx}$-$B$ curves [Fig. 1(b)].
The values of $\rho_{xx}$ is close to zero except
for the regions of the magnetization reversals
(Increase in  $\rho_{xx}$ near zero magnetic fields
is attributed to temperature increase upon the magnetic-field polarity reversal during the  field scan).
Figure 1(d) shows the changes of $\rho_{xx}$ and $\rho_{yx}$
when the Fermi energy $E_{\rm F}$ was tuned by the gate voltage $V_{\rm G}$
at $B$ = 0 T after a magnetic training.
A plateau of $\rho_{yx}$ = $h/e^2$ appears in the $\rho_{yx}$-$V_{\rm G}$ curve
accompanied by nearly vanishing resistivity plateau in the  $\rho_{xx}$-$V_{\rm G}$ curve
when $I_{\rm ac}$ = 1~nA (black curves).
With increasing  $I_{\rm ac}$,
the $V_{\rm G}$ range for the plateau  becomes narrower.
Above $I_{\rm ac}$ = 30~nA, $\rho_{yx}$ is largely deviated
from $h/e^2$ and $\rho_{xx}$ is lifted off from zero
even at the plateau center ($V_{\rm G}$ = 0~V)
which is assigned to the charge neutrality point (CNP).

\begin{figure}[tbp]
\centering
\includegraphics[width=8.0cm]{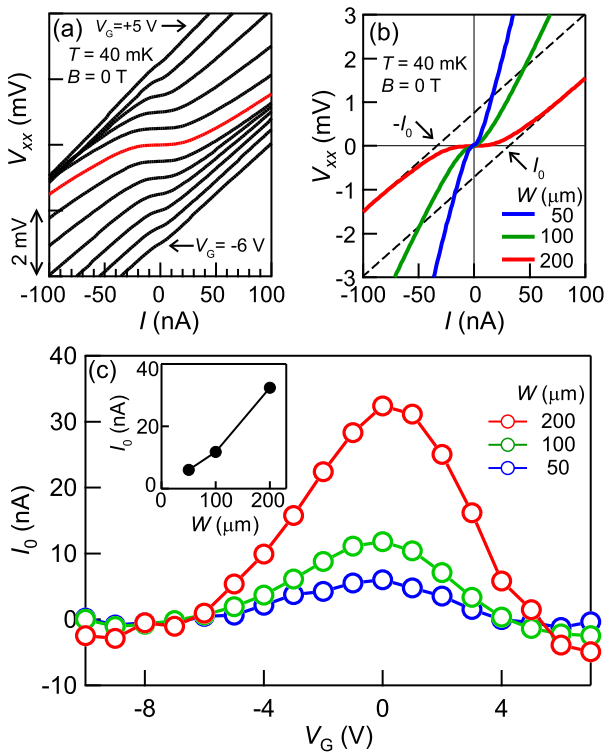}
\caption{\label{fig2}(color online)
	(a) $V_{xx}$-$I$ curves of the 200-$\mu$m-wide Hall bar
	at $T$ = 40~mK and  $B$ = 0 T under various gate voltages
	ranging from $V_{\rm G}$ = $-6$ V (bottom) 
     to $+5$ V (top) with an increment of 1~V.
	Traces are offset vertically by 0.5~mV for clarity.
	A trace for the CNP ($V_{\rm G} $ = 0 V) is highlighted in red color.
	(b)  $V_{xx}$-$I$ curves for the three Hall bars with different widths
	$W$ = 50 (blue), 100 (green), and 200 $\mu$m (red). 
	The curves were measured at the charge neutrality point
	of each Hall bar.
	(c) Variation of $I_0$ as a function of  $V_{\rm G}$.
	Data for the three Hall bars are shown.
	The inset shows the maximum value of $I_0$ for each Hall bar plotted
	as a function of the Hall-bar width $W$.
}
\end{figure}

The current-induced breakdown of the QAHE
can be seen more clearly in longitudinal voltage-current ($V_{xx}$-$I$)
characteristic curves obtained by the dc measurements.
Figure 2(a) shows the $V_{xx}$-$I$ curves for the 200-$\mu$m-wide Hall bar 
measured at $B$ = 0 T and $T$ = 40~mK under various gate voltages.
The $V_{xx}$-$I$ curves exhibit non-linear behaviors
with suppressed-$V_{xx}$ regions around $I$ = 0~nA.
As the Fermi energy is detuned from the CNP (a red curve), 
the $I$ range for the suppressed-$V_{xx}$ regime becomes narrower.
At the CNP, the value of $V_{xx}$ starts to increase
gradually when $I$ exceeds about 10~nA.
By further increasing $I$ above  50~nA,
the $V_{xx}$-$I$ curve approaches a straight line with a large slope.
Figure 2(b) shows the $V_{xx}$-$I$ curves for the three Hall bars with 
different widths under $V_{\rm G}$ tuned to each CNP.
The wider $I$ range for the suppressed-$V_{xx}$ regime
is observed in the  wider Hall bars.
We define a characteristic current $I_0$ for the crossover
from the suppressed-$V_{xx}$ regime to the linear regime
by a horizontal intercept of a line fitted to the $V_{xx}$-$I$ curve
 in the linear regime as shown by dashed lines in Fig. 2(b)
 (Detailed procedures are described in the Supplemental Material\cite{suppl}).
Figure 2(c) shows variation of $I_0$ as a function of $V_{\rm G}$.
The value of $I_0$, which takes a maximum at the CNP,
decreases as $V_{\rm G}$ is detuned from the CNP.
As shown in the inset of Fig. 2(c),
$I_0$ at the CNP is roughly proportional to the Hall-bar width $W$.

These behaviors of $I_0$
are similar to those of the critical current for the QHE breakdown\cite{Ebert, Okuno, Furlan, Komiyama},
which takes maximum near the center of a QHE plateau and
is proportional to the Hall-bar width.
The proportionality of $I_0$ to the Hall-bar width indicates 
that the Hall electric field, which appears between the edge channels 
across the Hall bar, is relevant to the crossover.
In the suppressed-$V_{xx}$ regime around $I$ = 0~nA,
the back scattering of electrons between the counter-propagating edge channels
are suppressed due to the magnetization-induced gap
of the top and bottom surfaces. 
As $I$ is increased,
the back scattering events start to take place with an assistance of the Hall electric field.
The values of $I_0$ are very small compared to a typical critical current
for the QHE breakdown\cite{Ebert, Okuno, Furlan}.
An abrupt jump in $V_{xx}$, which is usually observed
in the case of the QHE breakdown  at the critical current\cite{Ebert, Komiyama},
was not observed in the present study.
The abrupt jump in the QHE breakdown is attributed to thermal instability
of the QHE state under a large Hall electric field\cite{Komiyama}.
Unlike 2D electron systems in GaAs or graphene,
the surface state of the magnetic TI films have short scattering times.
We speculate that the short scattering time may prevent the surface electrons
from gaining a large kinetic energy to induce the abrupt jump.

\begin{figure}[tbp]
\centering
\includegraphics[width=8.6cm]{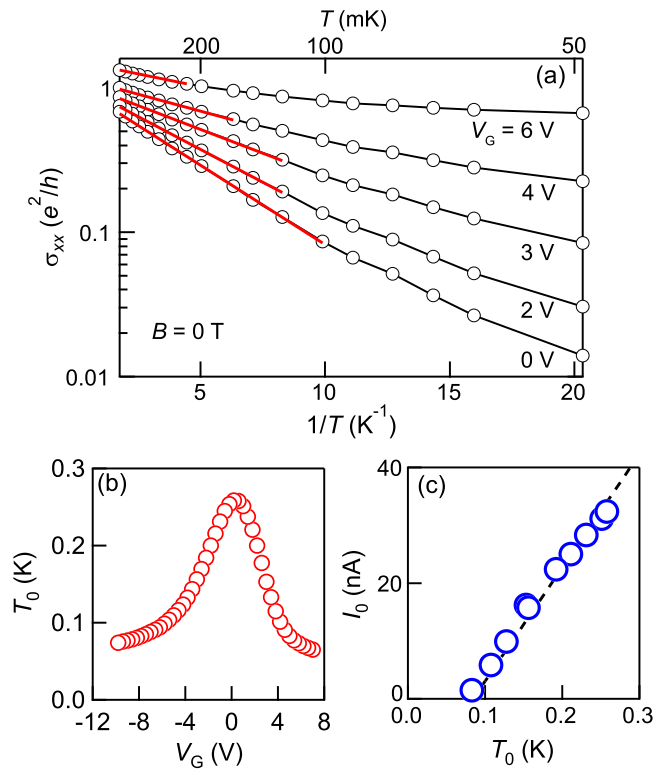}
\caption{\label{fig3}(color online)
	(a) Temperature dependence of the longitudinal conductivity $\sigma_{xx}$
	for the 200-$\mu$m-wide Hall bar	plotted as a function of $1/T$.
	Data for $V_{\rm G}$ =  0, 2, 3, 4, and 6 V  are shown from bottom to top.
	Red lines are the Arrhenius fitting results.
	(b) Thermal activation energy $T_0$ plotted as a function of $V_{\rm G}$.
	(c) Correlation between  $I_0$ and $T_0$ obtained at various $V_{\rm G}$.
	The dashed line shows a linear-fitting result. 
}
\end{figure}

\begin{figure}[tbp]
\centering
\includegraphics[width=8.6cm]{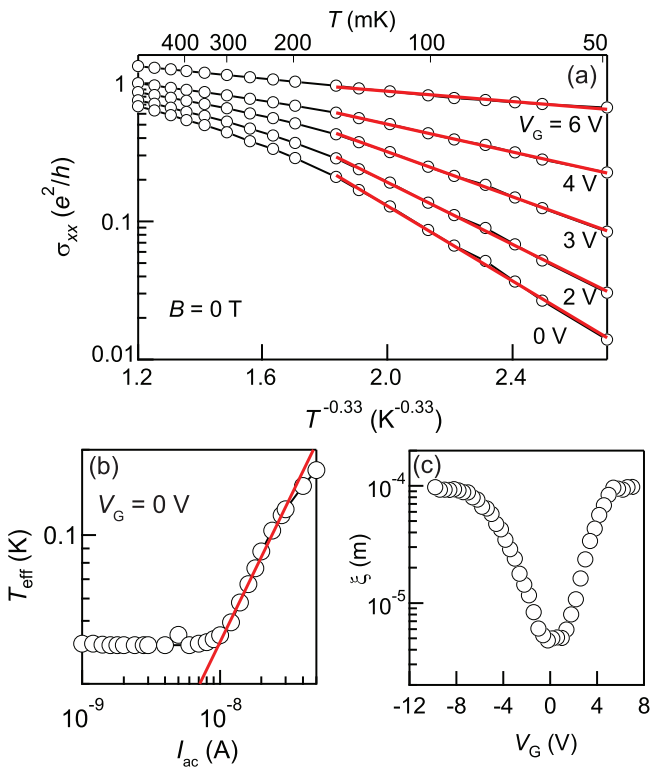}
\caption{\label{fig4}(color online)
	(a) Temperature dependence of $\sigma_{xx}$ for the 200-$\mu$m-wide Hall bar
	[the same data as in Fig. 3(a)]  plotted as as function of $T^{-0.33}$.
	(b) Effective temperature $T_{\rm eff}$  for $V_{\rm G}$ = 0 V 
	plotted as a function of $I_{\rm ac}$ using logarithmic scales.
	 $T_{\rm eff}$ is derived by comparing $\sigma_{xx} (I_{\rm ac})$ and $\sigma_{xx}(T)$
	point by point.
	The red line shows a linear-fitting result.
	(c) The $V_{\rm G}$ dependence of the localization length $\xi$
	estimated from the coefficients of the $T_{\rm eff}$-$I_{\rm ac}$ curves.
}
\end{figure}

To understand the relation between $I_0$ and 
the QAHE-observable temperature more quantitatively,
we compared $I_0$ to the thermal activation energy  
evaluated from the temperature dependence of $\sigma_{xx}$.
The values of $\sigma_{xx}$ are calculated from  $\rho_{xx}$ and $\rho_{yx}$ 
for each $V_{\rm G}$
(see Supplemental Material\cite{suppl} for details).
Figure 3(a)  shows temperature dependence of $\sigma_{xx}$
for the 200-$\mu$m-wide Hall bar  
plotted as a function of $1/T$ for several $V_{\rm G}$.
The thermal activation energy $T_0$ is  evaluated
by Arrhenius fitting to $\sigma_{xx} = \sigma_0 \exp (-T_0/T)$ for each $V_{\rm G}$.
Figure 3(b) shows the obtained $T_0$ as a function of  $V_{\rm G}$.
The peak value of $T_0$ at the CNP is about 260 mK, 
which is slightly larger than the value reported in the earlier work
on the Cr-doped (Bi, Sb)$_2$Te$_3$ films\cite{Bestwick}.
Figure 3(c) shows a correlation between $T_0$ and $I_0$
 obtained under various $V_{\rm G}$.
$I_0$ is almost proportional to $T_0$.
The linear relation between $I_0$ and $T_0$
is different from the case of the QHE breakdown\cite{Okuno}
where the critical current increases
with the cyclotron energy to the power of 3/2.
Although the origin of the different power is not clear for the present, 
the observed linearity  ensures that the stability of the QAHE against
current is expected to improve in accordance with the thermal stability.

Although $\sigma_{xx}$ follows the thermal activation-type 
temperature dependence at temperatures above 200 mK,
clear deviation from the fitted line can be seen at lower temperatures in Fig. 3(a).
The discrepancy suggests involvement of  
the other transport mechanism, such as variable range hopping (VRH).
Given that only the surface states can contribute to the electron transport,
the VRH transport would give a temperature dependence in the form of 
$\sigma_{xx}  = \sigma_1 \exp (T_1/T)^{1/(d+1)}$ with $d = 2$.
By replotting the same data in Fig. 3(a) as a function of $T^{-0.33}$,
we found that the data points at low temperatures reside on a straight line
for each $V_{\rm G}$[Fig. 4(a)].
We also analyzed the data using the 3D VRH model
and a VRH model with Coulomb interaction\cite{suppl}.
The experimental data and these models also show good agreements 
in limited ranges of temperature.
Although we cannot distinguish which  VRH model is the most preferable,
these results indicate that the VRH dominates electron transport at low temperatures.
The occurrence of the VRH transport means existence of a large number of in-gap states
within the magnetization-induced gap of the surface state.

The VRH  behavior allows us to evaluate the localization length $\xi$
by comparing the temperature dependence $\sigma(T)$ with 
the current dependence $\sigma(I_{\rm ac}$), 
following the earlier work  on the localization length analysis
in the QHE regime\cite{Furlan}.
[$\sigma(I_{\rm ac}$) is obtained from the similar measurement as in Fig.~1(d).
See Supplemental Material\cite{suppl} for the details.]
When an electric current $I_{\rm ac}$ is applied to 
a QAH conductor, a Hall voltage $V_{yx} = (h/e^2) I_{\rm ac}$ appears
between the edge channels.
The Hall voltage induces energy difference between the localized states,
assisting  electron hopping transport among them.
Assuming that the energy difference plays the same role
as temperature for the electron hopping,
the induced energy difference can be regarded as an effective temperature $T_{\rm eff}$.
The effective temperature under application of $I_{\rm ac}$ can be obtained
by comparing  $\sigma_{xx} (I_{\rm ac}) $ to $\sigma_{xx} (T)$ point by point.
Meanwhile, the averaged energy difference between the localized states
can be expressed by $e V_{yx} / (W / \xi )$,
supposing that the each localized state extends spatially over $\xi$.
Thus, using the relation $2 k_{\rm B} T_{\rm eff} =  e V_{yx} / (W / \xi )
 =  (h/e)(\xi/W) I_{\rm ac} $,
one can evaluate the localization length $\xi$ 
from the dependence of $T_{\rm eff}$ on $I_{\rm ac}$.
Note that this localization length analysis does not depend 
on the details of the VRH model\cite{suppl}.

Figure 4(b) shows the effective temperature $T_{\rm eff}$
as a function of $I_{\rm ac}$ for $V_{\rm G}$ = 0~V.
When $I_{\rm ac}$ is smaller than 10 nA,
the VRH transport is dominated by real temperatures
so that  $T_{\rm eff}$ does not depend on $I_{\rm ac}$.
But when $I_{\rm ac}$ exceeds 10 nA,
$T_{\rm eff}$ increases proportionally to $I_{\rm ac}$.
From the coefficient of the linear fitting result,
the localization length $\xi$ is evaluated to be 4.8 $\mu$m.
The similar analysis can be performed at each $V_{\rm G}$.
The obtained localization length is shown in Fig. 4(c) as a function of  $V_{\rm G}$.
The localization length taking a minimum value at around the CNP
 extends as the Fermi energy is deviated from the CNP.
The localization length of the present study is much longer
than the typical localization length reported in the QHE 
(about 100 nm for $\nu$ = 3 \cite{Furlan}) even at the CNP.
The long localization length indicates that the density of  states
 is still large in the middle of the magnetization-induced gap
and that the electrons hop around among these in-gap states.

Possible origins for the formation of the in-gap states
include  disordered Cr distribution.
A spectroscopy-imaging measurement\cite{Lee} revealed that 
the magnetization-induced  gap in Cr-doped (Bi, Sb)$_2$Te$_3$
reaches as large as 30 meV near the region where Cr density is high
while it decays as leaving away from the Cr-dense region.
Because $\sigma_{xx}$ in the transport measurement reflects
the smallest gap region connecting
the counter-propagating edge channels,
the energy gap of the Cr-sparse region probably contribute to the long $\xi$.
According to a recent theory\cite{Yue}, the magnetization-induced gap can survive
in the presence of such Cr disorder but the gap is largely reduced
compared to the case of uniform Cr distribution.
We speculate that the observed activation energy $T_0$
may correspond to the reduced magnetization-induced gap.
The observation of the VRH transport shows the presence of 
a number of localized states even in the reduced gap.
Other origins of disorder such as Te vacancy or anti-site defect
may also contribute to the formation of the in-gap states.

To summarize, we have demonstrated that the QAHE can be broken
when the current exceeds a certain value.
The characteristic current for the QAHE breakdown
is found to be proportional to the Hall-bar width.
These features have been discussed
 in comparison with the QHE breakdown.
The small characteristic current  is attributed
to the small magnetization induced gap containing a large number of in-gap states.
Reducing the number of the in-gap states is a key toward
a robust QAHE.

\acknowledgments
We thank T. Morimoto, N. Nagaosa, K. Yasuda, and M. Mogi for fruitful discussions. 
This research was supported by the Japan Society
for the Promotion of Science through the Funding Program for World-Leading
Innovative R \& D on Science and Technology (FIRST Program) on
“Quantum Science on Strong Correlation” initiated
by the Council for Science and Technology Policy,
JSPS/MEXT Grant-in-Aid for Scientific Research 
(No. 24224009, 24226002, and 15H05867),
and CREST, JST.


\end{document}